\documentclass[useAMS,usenatbib]{mn2e}
\usepackage{natbib}
\usepackage{times}
\usepackage[dvips]{hyperref}
\usepackage{graphicx}
\usepackage{amsfonts}
\usepackage{txfonts}
\usepackage{amssymb}
\def\calsio{Ca$_2$Al$_2$SiO$_7$\,}

\def\mum{$\mu$m}
\def\deg{^{\circ}}
\def\gs{\mathrel{\raise0.35ex\hbox{$\scriptstyle >$}\kern-0.6em
\lower0.40ex\hbox{{$\scriptstyle \sim$}}}}
\def\ls{\mathrel{\raise0.35ex\hbox{$\scriptstyle <$}\kern-0.6em
\lower0.40ex\hbox{{$\scriptstyle \sim$}}}}

\def\araa{Ann. Rev. Astron. \& Astrophys.}
\def\apj{Astrophys.\ J.}
\def\apjl{Astrophys.\ J.}
\def\apss{Astrophys. \& Space Science}
\def\aaps{Astron.\ \& Astrophys.\ Supp.\ Series}
\def\aj{Astron.\ J.}
\def\mnras{Mon.\ Not.\ R.\ Astron.\ Soc.}
\def\nat{Nature}
\def\pasp{Pub.\ Astron.\ Soc.\ Pacific}

\def\aap{Astron. \& Astrophys.}

\title[Resolving the obscuring torus in NGC 1068 with the power of infrared interferometry]{
Resolving the obscuring torus in NGC 1068 with the power of infrared interferometry: Revealing the
inner funnel of dust}
\author[Raban et al.]{David Raban$^{1}$\thanks{E-mail:
raban@strw.leidenuniv.nl}, Walter Jaffe$^{1}$, Huub R\"{o}ttgering$^{1}$, Klaus  Meisenheimer$^{2}$
\newauthor and Konrad  R. W. Tristram$^{2,3}$\\\
$^{1}$Sterrewacht Leiden, Leiden university, Niels-Bohr-Weg 2, 2300 CA  Leiden, The Netherlands\\
$^{2}$Max-Planck-Institut f\"{u}r Astronomie, K\"{o}nigstuhl 17, D-69117 Heidelberg, Germany\\
$^{3}$Max-Planck-Institut f\"{u}r Radioastronomie, Auf dem H\"{u}gel 16, D-53121 Bonn, Germany}
\begin{document}
\bibliographystyle{mn2e}
\date{}
\pagerange{\pageref{firstpage}--\pageref{lastpage}} \pubyear{2008}
\maketitle
\label{firstpage}
\begin{abstract}
 We present new interferometric data obtained
with MIDI (MID infrared Interferometric instrument) for the Seyfert II galaxy NGC~1068,  
with an extensive coverage  of sixteen \textit{uv} points.
These observations resolve the nuclear mid-infrared
 emission from   NGC~1068 in unprecedented detail
with a maximum resolution of 7 mas.
 For the first time, sufficient \textit{uv} points have been obtained, allowing us to generate an image
of the source using maximum entropy image reconstruction. The features of the image
are similar to those obtained by modelling.
 We find that the mid-infrared emission can be represented
by two components, each with a Gaussian brightness distribution. The first,
identified as the inner funnel of the obscuring torus, is
hot ($\sim$800K),  1.35 parsec long, and 0.45 parsec thick in FWHM 
 at a PA=$-42^\circ$ (from north to east).
 It has an absorption profile different than standard interstellar dust
and with evidence for clumpiness.  The second component is $3\times 4$ pc in FWHM 
with T=$\sim$300K, and we identify it with the cooler body of the torus.
The compact component  is tilted by $\sim45^\circ$ with respect to the radio jet and has
similar size and orientation to the observed water maser distribution.
We show how the dust distribution  relates to other observables within a 
few parsecs of the core of the galaxy such as the nuclear masers, the radio jet, 
and the ionization cone.
We compare our findings to a similar study of the Circinus galaxy and other relevant studies.
 Our  findings shed new light on the relation between the different parsec-scale components in
 NGC\,1068 and the  obscuring torus.
\end{abstract}
\begin{keywords}
techniques: interferometric -- galaxies:
Seyfert --  galaxies(individual):NGC1068-- infrared: galaxies
\end{keywords}
\section{Introduction}
The AGN unification model \citep{Antonucci93, Urry95} explains the difference 
between type II Seyferts, which show
only narrow emission lines,  and type I, which show both narrow and broad emission lines,
by stipulating that a torus-like structure surrounds the central engine and accretion disk.
Thus, when oriented edge-on, the torus blocks the subparsec-sized broad emission
line region making the object appear as a type II, 
while harbouring an unseen type I nucleus. Symmetry and angular
momentum considerations suggest that the torus and the accretion disk are perpendicular 
to the bipolar jet, whose orientation is likely to be coupled closely  
to that of the accreted matter.
The relative number of Seyfert I/II  demands that the 
torus be geometrically thick \citep{Osterbrock88}, although thick 
structures orbiting compact objects will
quickly lose their height and collapse into a thin disk. In order to maintain its
stable torus structure, the random velocities of the dust clouds 
need to be of an order similar to
the orbital speed, or a few hundred km/s.  Since colliding dust clouds are destroyed at
relative speeds as low as a few meters per second, the  dust clouds are not expected to
last very long \citep{Krolik88}.
To overcome the difficulty of maintaining its inflated state different dust 
configuration other than a torus have been
introduced, such as the warped disk of \citet{Sanders89}. Alternatively, the dust has been
described as a hydrodynamically driven outflow  \citep{Kartje96}.

NGC~1068 is considered to be the prototype Seyfert II galaxy, 
where the central source is obscured by dust.  
Its relatively small distance of 14.4 Mpc and high mid-infrared
flux make the object  ideally suited for the study of the
nucleus and obscuring dust. Previous MIDI observations of NGC~1068 
revealed warm ($320$ K) silicate dust in a structure 2.1 parsecs 
thick and 3.4 parsecs in diameter,
surrounding a small, hot ($800$K) component whose shape and
orientation could not be determined in detail  \citep{Jaffe04a} (hereafter J04).
These observations were, together with those of \cite{Wittkowski04} (W04), 
the first to spatially resolve direct emission from the putative torus, 
and the first to show that a torus-like structure is indeed present in NGC~1068.  

The mid-infrared emission from the central region of NGC~1068, 
as seen by the largest single dish
telescopes, is composed of an unresolved core plus extended emission. Deconvolved maps at
12\mum\, taken with VISIR (Imager and  Spectrometer in the InfraRed at the Very Large
Telescope)  by \cite{Galliano05} reveal a set of discrete mid-infrared sources: 
seven in the north-eastern quadrant and five in the south-eastern quadrant.
The central source, observed with the 10m Keck telescope \citep{Bock00},  is extended
by $\sim$1$"$ in the north-south direction and is unresolved in the east-west direction.
About 2/3 of its flux can be ascribed to a core structure which 
is itself elongated north-south
in a “tongue”-shaped structure and according to \cite{Bock00} does not show a distinct
unresolved core.
Recent 12.8\mum\, speckle images taken with VISIR in BURST mode by \cite{Poncelet07}
identify two major sources of emission at 12.8\mum: a compact source ($<85$\,mas) and an
elliptical source of size ($<140$) mas $\times$ 1187 mas and PA$\sim4\deg$.
It is the unresolved compact source which is associated with the dusty 
torus and is the subject of this study.
 Other parsec scale components in NGC~1068  that are  related to the torus are the
compact H$_2$O masers , appearing on the sky as a
set of linear spots (PA=$-45^\circ$) that are misaligned with the jet  and span a velocity range
 of $~600$ km/s with a  sub-Keplerian velocity profile ($v\propto r^{-0.3}$). The kinematics of the maser spots
indicate that the masers are located in a rotating disk with inner radius of $\sim$0.65 pc
and outer radius of $\sim$1.1 pc \citep*{Gallimore01}.
This disk traces the outer,  colder part of the accretion
disk where conditions allow for the formation of water masers.
It was  proposed earlier that these
masers trace warm molecular dust \citep{Claussen86}, a conclusion supported by the
observations presented here which reveal the dusty torus to be at the outer edge
of the maser disk.   VLBA 5 and 8.4GHz radio continuum images show
a parsec-sized structure with a major axis at PA$\simeq$-75$^\circ$, most likely indicating
free-free emission from hot (T=$10^4 - 10^5$ $\,$K) ionized gas \citep{Gallimore04}.
According to the unification model, toroidal obscuration by dust is also responsible for the
conical shape of the narrow-line region. The ionization cone in NGC~1068 as seen in HST
images is  centred around PA=$10\deg$, while modelling of HST spectra,
based on the kinematics of the gas  indicate  the ionization cone with an opening angle of $80^\circ$
centred around PA=$30^\circ$ \citep{Das06}, roughly perpendicular to the maser spots.

\section{Observations, \textit{uv} coverage and  data reduction}\label{sec:obs}
MIDI is the mid-infrared interferometer located on Cerro Paranal in northern Chile and
 operated by the European Southern Observatory (ESO).  MIDI functions  as a classical
Michelson interferometer, combining the light from two 8-meter unit telescopes (UTs).
For a detailed description of the instrument see \citet{Leinert03}.
The main data product of MIDI is the correlated flux, which can be explained as
the spectrum of the source at a certain spatial resolution projected
perpendicular to the baseline, that is to the projected separation of the two
telescopes.
\begin{figure}
\begin{centering}
\includegraphics[width=7cm]{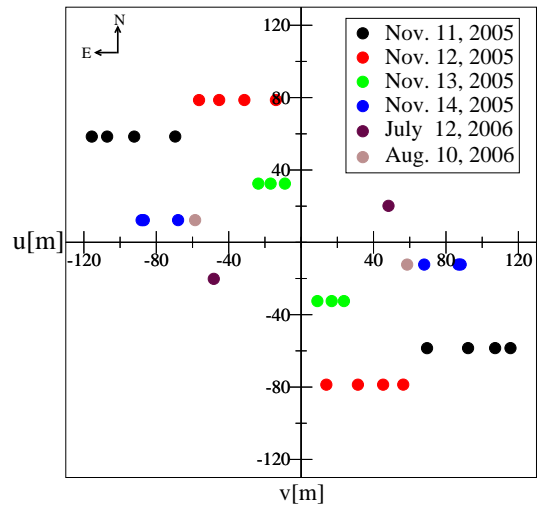}
\caption{\textit{uv} coverage [m] for NGC1068 and colour coded by date. Due to the $\sim$0.0 declination
of NGC1068, the \textit{uv}  tracks are parallel to the u-axis.  \textit{uv} coords [u,v]  are complex conjugates of
uv coords [-u,-v], and both are plotted since they are indistinguishable. \label{fig:uv}}
\end{centering}
\end{figure}
A total of 16 interferometric observations were taken together with a single-dish total flux
spectrum for each one.  Table \ref{tab:obslog} contains  the observation log and instrument settings.
\begin{table}
 \caption{ Log of the observations dates, unit telescopes used  and baseline
configuration in  polar (D, PA) coordinates projected on sky.  All observations
were done in GRISM mode with a spectral resolution of $\lambda/\Delta\lambda=230$.
The same calibrator, HD10380, was used for all observations with identical settings.
 Chopping with $f=1$Hz, and throw amplitude of 15'' at  an angle $\alpha=0^\circ$ was
applied  during photometric measurement and acquisition.
Fringes were tracked in ''offset tracking'' mode. The spatial resolution  specified is calculated
by $\theta=\lambda/(2.2B)$ where B is the projected baseline separation and $\lambda=10\mu m$.
 \label{tab:obslog}}
\begin{tabular}{@{}lccccc}
 \hline
       &\textbf{Date} &\textbf{UT's}&\textbf{D}& \textbf{PA}& \textbf{resolution} \\
       &  &   &    [m]&deg&[mas] \\
\hline
\#1 &Nov. 11, 2005       &    UT1-UT4   & 90.8      &  50.0        & 10.3\\
\#2 &                           &          ''         & 109.2    &  57.6      &  8.6 \\
\#3 &                           &      "             & 122.      &  61.4      &  7.7\\
 \#4 &                          &  ''                 & 129.6    &  63.2      &  7.2\\
\#5 &Nov. 12, 2005       &  UT1-UT3    &  79.9      &  10.0          & 11.7\\
\#6 &                           &         ''          & 84.7      &  21.7       &  11\\
\#7 &                           &         ''          & 90.8     &  29.9       & 10.3\\
\#8 &                           &       ''            & 96.8     &  35.6     & 9.7\\
\#9 &Nov. 13, 2005       &   UT2-UT3    & 33.6     &  15.5       & 27.9\\
\#10 &                         &       ''            & 36.6     &  27.5       & 25.6\\
\#11 &                         &       ''            & 40.15   &  36.1       & 23.4\\
\#12 &Nov. 14, 2005     &   UT2-UT4   & 69.0     &  79.8        & 13.6\\
\#13 &                         &      ''            & 89.98   &   82.1        & 10.4\\
\#14  &                        &         ''          & 87.76   &   82.0          & 10.7\\
\#15 & July 12, 2006     &   UT3-UT4    & 52.3     & -67.3       &  17.9\\
\#16 & Aug. 10, 2006    &   UT2-UT4    & 59.82   &  78.2       & 15.6\\
\hline  
\\
\end{tabular}
\end{table}
A map of the \textit{uv} coverage is shown in Fig. \ref{fig:uv}.  Due to the near  zero declination
of NGC\,1068 the tracks are parallel to the u-axis, causing each observation to differ both by
baseline length and position angle from its neighbours. The distribution of the tracks
is concentrated in the second/fourth  quarter of the \textit{uv} plane, with only one observation on a
 perpendicular baseline, reflecting  the fixed physical placement of the VLTI unit  telescopes.
After each interferometric observation, a total flux measurement (i.e. single-dish spectrum) 
is taken from each  telescope independently, and used to determine the visibility, i.e. the
correlated interferometric flux
divided by the geometric mean of the single dish fluxes.
 Only one such observation was rejected due to bad seeing, and the remaining 15
 fluxes were averaged together (Fig. \ref{fig:totflux}).
 A typical MIDI observation is recorded
in 8000 frames, with an integration time  of 0.036 seconds for each frame. We examined
the total flux recorded on the detector per frame, and rejected frames where the flux
was unstable.
Data reduction was done by coherent visibility estimation with the Expert Work Station (EWS)
 package written  in Leiden. See \citet{Jaffe04b} for  a detailed description
 of the coherent visibility estimation method.
\subsection{Calibration}                                                        
Calibration of the interferometric and photometric data was done using HD10380 as the
calibrator star for all  observations.  Each night's observation started with  HD10380, and it was
observed between observations of  NGC~1068. Thus, we obtained a calibration measurement for
each \textit{uv} point of NGC~1068. The calibrator's data reduction was done using the same parameters as
for NGC~1068 and went through the same tests for consistency as for the science target.
Only one calibrator observation was rejected due to bad seeing.
Calibration measurements taken just before and after NGC~1068
 were compared and, if different, were averaged.
For an unresolved point source, the correlated flux should equal the total flux. Assuming this,
we can calibrate the correlated fluxes using the calibrators, without a need to use the photometry
measurements, which are less reliable due to the strong atmospheric background.
The calibrated fluxes are computed by dividing the correlated fluxes of the target
by those of the calibrator,  and then multiplying by the known flux of the calibrator. For HD 10380 we
used the spectral template of \citet{Cohen99}. The calibrated photometric
 fluxes were produced by the same procedure for the photometric data.
 \subsection{Correlated flux. vs. visibility}
Optical/infrared interferometric data is often presented as visibility data, a dimensionless
 quantity, which measures the fraction of flux that is resolved for a  given baseline.
 Here we prefer to use the correlated fluxes, the actual flux that
the interferometer measures for a given baseline, as is  common practice in radio interferometry.
  Determining the visibility, $V$,  directly from the contrast of the
interferometric fringes ($V=(I_{max}-I_{min})/(I_{max}+I_{min})$) is not practical
since the amount of flux reaching the beam combiner from the
two telescopes is generally not equal and therefore affects the contrast of the fringes.
Given an isolated source, the visibilities equal the correlated
fluxes divided by the total flux of the source.
 If, however, the source is embedded in emission on larger scales
(as seems to be the case for NGC~1068), then the total flux  determined by a larger field of view
is contaminated by the large scale emission making it difficult to interpret.
The correlated fluxes also free us from handling the strong atmospheric background,
 which is not correlated and is removed easily in the data reduction process.
\section{Results}\label{sec:results}                                                   
\subsection{The total flux}
\begin{figure}
\includegraphics[width=7cm]{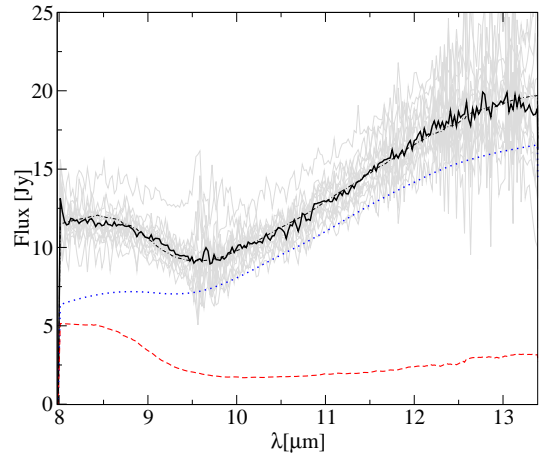}
\caption{Total (single-dish) flux.  All 15 individual  spectra are in grey with the mean given in
black. The dash-dotted line is the model fit described in section \ref{sec:bb2}, which is
composed of two flux components: component 1 (red-dashed) and component 2 (blue-dotted).
Note the atmospheric O$_3$ feature at 9.7\mum.
\label{fig:totflux}}
\end{figure}

The total flux of NGC~1068 is shown in Fig. \ref{fig:totflux}, as measured by 
the average of two  8-meter UTs, each  with a beam size of 250 mas .
It is in  rough agreement with previously published spectra from 8m class telescopes.
The only spectral feature seen in the total flux is the 9.7\mum\, silicate feature, in absorption.
We do not detect the very weak [SIV] emission line reported by \cite{Rhee06} and \cite{Mason06}
with similar apertures. Space based instruments show PAH features at $\lambda=7.6,9\,\textrm{and}\,10.9$\mum\, 
\citep{Sturm00}. These features are found to occur at larger distances from the nucleus.
\begin{figure}
\includegraphics[width=7cm]{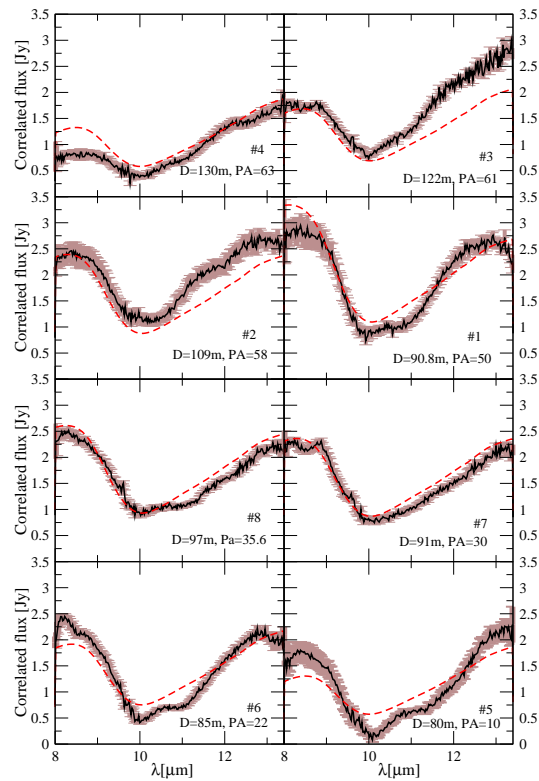}
\caption{Correlated fluxes \#1-8 (solid black) and the model fit from section \ref{sec:bb2}
 (red-dashed). The panels are sorted according to baseline length and the unit telescopes used, and
 are numbered according to the number in the observation log (Table \ref{tab:obslog}).
  \label{fig:corrflux1}}
\end{figure}
 
\begin{figure}
\includegraphics[width=7cm]{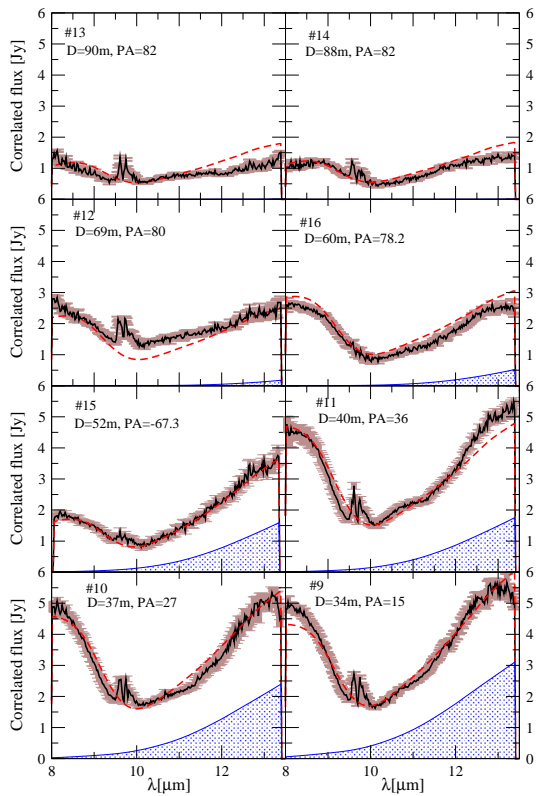}
\caption{Correlated fluxes \#9-16 (solid black) and the model fit described in section \ref{sec:bb2} (red-dashed).
  The blue filled curves show  the contribution of component 2 to the correlated fluxes.
The `bump' at 9.7\mum\, visible on fluxes \#9-14,16 is due to the atmospheric O$_3$ feature.
The panels are sorted according to baseline length and the unit telescopes used, and
 are numbered according to the number in the observation log (Table \ref{tab:obslog}).
\label{fig:corrflux2}}
\end{figure}
\subsection{The correlated (interferometric) fluxes }
The correlated fluxes, shown in Figures  \ref{fig:corrflux1}\&\ref{fig:corrflux2}  display large
 variations in the shape and depth
of the silicate feature centred at 9.7\mum, the only spectral feature present in the data.
 No traces of PAH  molecules or the 12.9\mum\, Ne line are seen.  The clear relation seen
in the data between the baseline length and the correlated flux  shows that the source is 
 resolved on every baseline.
The `bump' around 9.7\mum\, seen in baselines \#9-14,16 (Figures \ref{fig:corrflux1}\&\ref{fig:corrflux2} ) is due to the
atmospheric ozone feature, centred 9.7\mum.
\subsection{Discussion of the results}\label{results_discuss}
The correlated fluxes  in fact sample  the SED of
NGC~1068 about a range  of different orientations and spatial resolutions. The structure of the
spectra, and in particular the 9.7\mum\,  silicate feature,  varies considerably with 
baselines, and is quite different from the total flux.  
The striking differences between the correlated and total fluxes demonstrate  how
 inadequate the total flux is  as the sole means to constrain torus models. Indeed, numerous
authors have attempted to fit tori models to the  SED of NGC~1068,
and many different models fit the SED equally well, as was first demonstrated by \cite{Galliano03}.
 
The depth of the silicate feature in the correlated fluxes is about 0.6 of the continuum
level (as defined by the average of the flux at 8 and 13\mum) and is independent of
baseline length. This suggests that the emitting medium is compact and  is located in 
a region much smaller than the absorbing medium. 
 For the total flux, the silicate feature depth is smaller, $\sim0.3$.  
The total flux  can be successfully fitted with a combination of two grey body spectra,
a warm ($\sim$300K) component, plus a hot ($\sim$800K) compact
component, both behind uniform absorption screens as discussed in \S\ref{sec:bb2}.
In general we find the total flux to be easily reproducible  by our models described in
\S\ref{sec:bb2} for a wide range of parameters.
\subsection*{A note on phases and symmetry}
Apart from the correlated and total fluxes, one can also recover the differential phase  $\phi_{diff}(k)$
of the source from MIDI data. The differential phase is related to the
source's  ‘true’ phase, $\phi(k)$, by $\phi_{diff}(k)=\phi(k)-C\times k$, where
 $k=2\pi/\lambda$ is the wave number and $C$ is an unknown constant. 
Thus, any linear dependence of $\phi$ on $k$ is removed from the differential 
phase in the data reduction process, and cannot be recovered.
This missing information in the differential
phase makes it very challenging to include it in our models. We have therefore
postponed treatment of the differential phase in NGC~1068 to a second paper.
Without the phase information, only models or images which possess
inversion symmetry can be applied to the data. Throughout this paper, we assume a
symmetric flux distribution. This is an instrumental limitation and we do not claim that
 the true source brightness possesses such symmetry.
\section{Modelling}\label{sec:modeling}
As interferometric data is obtained in Fourier space,  interpreting the data is not as
 straightforward as in direct imaging or spectroscopy. In infrared interferometry, where only a small number
of the absolute value of the Fourier components can be measured,  the results are often model dependent and difficult to
  interpret unambiguously. The effective resolution for each \textit{uv} point is determined
 by $\lambda/ B$, the projected separation  between the telescopes expressed in number of wavelengths.
This relation causes the spectral features of
the source, which strongly depend on $\lambda$, to become entangled with the spatial flux
distribution as observed by telescopes of separation $B$.
 We have attempted to disentangle the
two effects by using two different modelling approaches. First, we treat the flux distribution
as coming from two Gaussian components, each with a fixed size and orientation. Each component
is assumed to be a grey body with a fixed temperature and situated behind a uniform absorption screen
(\S\ref{sec:bb2}).
The model results show us that at 8\mum\, only one component contributes to the correlated flux.
Therefore, we chose to attempt to reconstruct an image at this wavelength, which is shown 
and discussed in \S\ref{sec:ME}. The image 
agrees well with the modelling results. 
In our third approach (\S\ref{sec:onegauss}) we  look at each wavelength
 independently and fit a single Gaussian component to each, assuming no relation between the
different wavelengths in the data,
and assuming all the observed structures in the correlated fluxes arise from the
different source sizes and orientation at each wavelength  (channel by channel fitting).
 A good understanding of the observed properties of NGC
1068 can  then be  achieved by combining the results from the different procedures.
\subsection{Two grey body model}\label{sec:bb2}
The first approach to disentangling structure and spectrum is to assume that
changes to the correlated fluxes with wavelength are solely due to  spectral effects,
i.e. a grey body emission undergoing absorption while keeping the spatial parameters
constant with wavelength. In this model we treat the infrared emission as coming from two
Gaussian grey body components   of a fixed size and orientation, each one behind a uniform
 absorption screen. Two  is the minimal number of components necessary  to account well for the data,
while the addition of more components does not improve our fit.
The correlated flux  of each component is given by:
\begin{equation}\label{eq:a}
F_{corr}(\lambda,u,v)=\eta BB(\lambda) V(u/\lambda,v/\lambda) e^{-\tau\cdot abs}
\end{equation}
where BB($_{\lambda}$) is the  emission from a black body  of temperature $T$; $V$ is the visibility
of a Gaussian component whose major axis, minor  axis, and  PA are to be fitted;
$abs$ is the absorption curve of a chosen mineral or a combination of a few minerals 
as described  below.
The coefficient $\eta$ is the grey body scaling factor and  has  a value  0$<\eta<$1, independent
of wavelength.
\begin{figure}
\includegraphics[width=7cm]{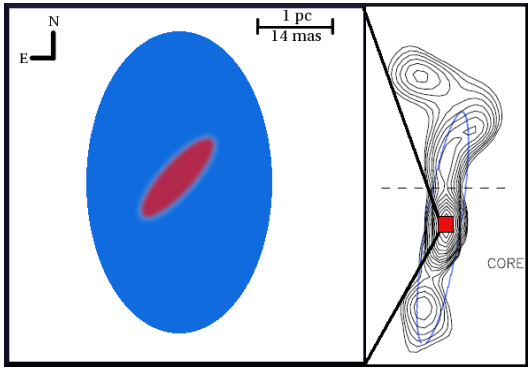}
\caption{Comparison between  the two components in our model and the 12.5\mum\, image of
\citet{Bock00}, taken with the 10m Keck telescope. The two components are plotted
symmetrically, yet their relative position cannot be determined from the correlated fluxes.
 \label{fig:compare}}
\end{figure}
After computing the Fourier amplitudes corresponding to each \textit{uv} point in the model, the
two sets of amplitudes (one for each component) are then combined to produce the final
correlated flux to be compared  to the data:
\begin{equation}\label{eq:b}
F_{corr}=\vert F_{corr 1}+e^{2\pi i (ul+vm)}F_{corr 2}\vert
\end{equation}
The parameters $l$, $m$  are for the two components to be moved with respect to each other in the
plane of the sky, parallel to the RA and DEC directions, respectively.
 In practice we find that the data does not provide constraints for
the relative positions $l,m$ (see also \S\ref{sec:symm}) and we have set them to zero (concentric components).
Each component in the model has six parameters, making in twelve in total: the major and minor FWHM,
the position angle $\phi$, the BB temperature T, the optical depth $\tau$, and the scaling
factor $\eta$.
The main features of the data can be 
reconstructed from values of the correlated flux at about six wavelengths, in the 
sense that if the model fits the data well at those wavelengths it will fit the rest well. 
We can then estimate that these 12 parameters are in fact fitting $6\times16=96$
independent data points.
The total flux is treated here as an extra \textit{uv} point to be fitted, and is computed by setting
the $u$ and $v$ coordinates in equation \ref{eq:b} to zero for each component.
In practice we find that the total flux is easily reproduced in many of our models
and for a different range of parameters, and may serve as an upper limit
to the flux and size of each component.
The most difficult part in our model is to determine the absorption curves in equation \ref{eq:a}. 
We have selected several dust absorption templates, including standard galactic dust  as
 observed towards the centre of the Milky Way  \citep{kemper04}, as well as \calsio, which  was
  found to best fit the previous MIDI data  (J04, \cite{speck00}).  Our aim here is not to
 unambiguously determine the chemical composition of the dust, which is not possible 
   to do from the data,   but rather to characterise the spectral template which will fit 
   the data best, and to compare it with other known dust mineral templates.
   In this respect the template is 
 any simple  function $f(\lambda)$ which,
  when inserted into equation \ref{eq:a}, will  provide the best fit for the data.  
To achieve this, we first fit a combination of the minerals listed above and shown in Figure
 \ref{fig:absorbers}.  The best-fit combination was improved on by allowing the value
  of $\tau$ at several “key”  wavelengths (i.e. 8, 9, 9.7, 11.5 and 13\mum) to vary and 
  then interpolated between the new values using cubic spline. This method is  able to 
  mimic a typical absorption template.
 The resulting template, which we designate  `fitted composite', along with several 
 other dust templates are plotted in Fig. \ref{fig:absorbers} while the model parameters
   are given in Table. \ref{tab:fit}. We label the two components in our model as  
  components 1 and 2.  Figures \ref{fig:corrflux1}\ and \ref{fig:corrflux2}  plot the best 
  fit model against the correlated fluxes, while Figure \ref{fig:totflux} plots the model's
   fit to the total flux  plus the contribution from the individual components.
The fitting was done by finding the least-squares solution using the  Levenberg-Marquardt technique.
The general trend is for the quality of the fit to become worse with increasing baseline length, 
as one might expect from the lower spatial resolution of a short-baseline observation. 
In addition, component 2 only contributes to the correlated fluxes taken with shorter baselines, 
effectively increasing the number of fit parameters.
Some of the features in the correlated fluxes, in particular the `step' seen most clearly
 in correlated fluxes \#4 and 5 at 10.5\mum, cannot be reproduced by the simple
 model we propose. Our model  will always generate smooth fluxes due to the smooth
  absorption templates we use, which not contain any such `steps'. 
 Furthermore, introducing more complicated templates will not help to fit these unsmooth features
 since any change made to the templates will affect each correlated flux, while these ‘unsmooth’
 features only appear in some of the correlated fluxes.  
  
  Seven out of sixteen modelled correlated fluxes
 (\#7,8,9,10,11,15,16) are in excellent agreement 
with the data while the rest of the modelled fluxes are mostly in 
good to perfect agreement with the data
either for $\lambda<$10\mum\, or for $\lambda>$10\mum.
The two components are illustrated in Figure \ref{fig:compare}.
Given the range of spatial resolutions of the different baselines (spanning a factor $\sim$4 in length),
 the complex behaviour of the correlated fluxes  and the striking simplicity of the
model, it provides a surprisingly good fit to the data.
\begin{figure}
\includegraphics[width=7cm]{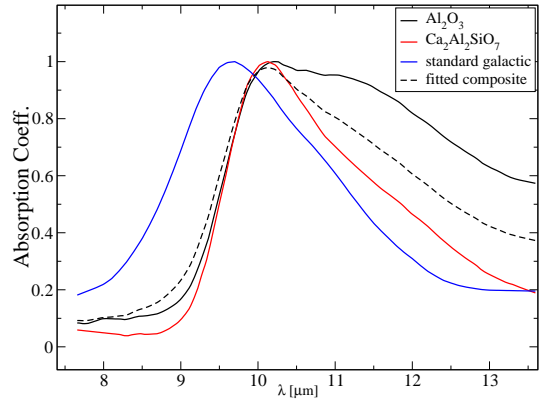}
\caption{ A comparison between the absorption template used in fitting the data
(dashed) and the other spectral templates for dust particles commonly found in astrophysical environments.
The latter includes that of dust  seen toward the centre of the Milky Way.  These templates only include
line absorption and do not include the continuum absorption.
\label{fig:absorbers}}
\end{figure}
\begin{table}
\caption{Parameter fits for the two grey-body model of \S\ref{sec:bb2}.
The values here are  based on the results of different  models that differ
in the chemical composition of the dust.   
To determine the errors we fixed the value of all but one parameter at a time and 
noted the range of values resulting in significant changes to the model. \newline
$^*$The temperature and scaling factor, $\eta$, are not independent.
Temperatures as high as 1500K can be fitted with a
low scaling factor, $\eta\sim$0.05, although the resulting fit is not as good.}\label{tab:fit}
\begin{tabular}{|c|c|c|c|}
\hline Parameter          & value       & error               & units \\
\hline
                           & \textbf{Component 1} &              &\\
\hline
FWHM major      & 20                &$\pm$ 3    & mas \\
FWHM minor      & 6.4               &$\pm$1   & mas \\
 $\phi$              &  -42             &$\pm$ 2  & degrees \\
 T                      & 800$^*$       &$\pm$  150         &K \\
$\tau$               & 1.9               &$\pm$  0.5         &  \\
$\eta$            & 0.25$^*$      &$\pm$  0.07        &  \\
\hline
                         & \textbf{Component 2} &              &\\
\hline                         
FWHM major      & 56.5            &$\pm$  5    & mas \\
FWHM minor      &  42.4           &$\pm$5           &mas\\
$\phi$               &  0               & $^{+70}_{-20}$          & degrees \\
T                       & 290            &$\pm$  10         & K \\
$\tau$               & 0.42           &$\pm$  0.2          &  \\
$\eta$            & 0.64           &$\pm$  0.15           &\\ \hline
\end{tabular}
\end{table}
\subsubsection{Symmetry issues}\label{sec:symm}
As stated before, only phaseless and therefore symmetric models were applied to the data.
The lack of an absolute phase also implies a loss of astrometric information. We cannot, for example, determine
the position of the mid-infrared emission in relation to other known components on similar scales for NGC~1068.
In principle,  we could constrain the relative position of our two model components by means
of the $l,m$ parameters in equation \ref{eq:b}. If our two components are not symmetric with respect to
each other, the resulting phase difference will also affect the sum of the Fourier amplitudes of
both components,
i.e. the observed correlated fluxes.  In practice, we find that  we do not have enough information
on component 2 in order to determine its relative position. Models with offsets   of up to a few tens of mas
are virtually indistinguishable from models with zero offsets.  In order to better constrain the relative positions,
more observations on shorter baselines are needed, as component 2 is mostly over-resolved in our
baseline sample. We do know from the differential phase that the mid-infrared emission is asymmetrical. However,
we cannot tell whether this asymmetry is due to the geometry of the dust clouds or the result of
asymmetrical absorption in the torus.
\subsubsection{The infrared SED}
The model used here is specifically designed to explain the MIDI data in the wavelength 
range of 8-13\mum. It is not a full radiative transfer model, and therefore cannot 
be used to predict or account for the entire infrared SED of NGC\,1068.  
\subsection{Maximum entropy imaging}\label{sec:ME}
\begin{figure}
\includegraphics[width=7cm]{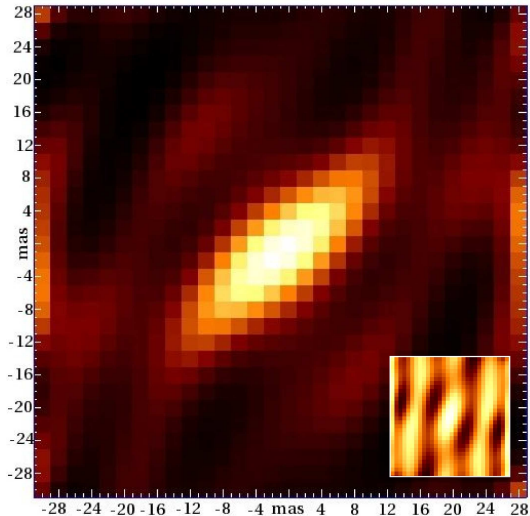}
\caption{Maximum entropy reconstruction at 8\mum. Image size is $30\times30$ pixels, with 1pix=2 mas.
Colour scale is linear. The extended blobs are artifacts caused by the low \textit{uv} coverage. Gaussian
fitting to this image
measures it to be 7.7 / 21 mas in FWHM with PA=$-46\deg$, very close values to
 the grey body model results,
and in agreement with the results of the one Gaussian fitting  at 8\mum.
The `dirty map' is shown in the bottom left corner.
\label{fig:mem}}
\end{figure}
The fits of the two blackbodies presented in \S\ref{sec:bb2} are not perfect fits to the data.
Attempting to improve them, we explored different perturbations in the Gaussian distribution
and although some of them can make the fit better, they  are still ad-hoc. Finally we decided
to reconstruct an image using maximum entropy (ME) methods, which guarantees that the resulting
image will be the most statistically probable reconstruction given the information in the data.
To our knowledge, this is the first time such a method has been used with infrared interferometric data.
The ME reconstruction is based on the general algorithm by Skilling \& Bryan (1984) and modified
to the specific needs of dealing with MIDI data.  The code was tested on simulated data
with similar S/N as the original data  and using the same \textit{uv} coverage of NGC~1068 (Fig. \ref{fig:uv}).
In general, the code is able to accurately reproduce Gaussian brightness distributions or similarly simple
shapes without sharp edges. Attempting to reconstruct multiple component images (such as a binary)
have not been successful for this \textit{uv} coverage.
We have chosen to reconstruct the image at 8\mum,  since in that wavelength a single component
fits the data well and the silicate absorption does not come into play.
The resulting image is displayed  and discussed in Fig. \ref{fig:mem}.
The ME method is useful since it is not model-dependent and no prior assumptions about the source
are needed to make use of it. Although the ME image details the basic properties of the source,
and can provide a good starting point for any model, it does not provide a more detailed picture
than parametric  models which enable us to probe the wavelength dependence of the data as well
as its brightness distribution. Nevertheless, the ME image does fit the data set at 8\mum\,  perfectly.
Since the code, by definition,  always prefers to get rid of complicated structures, we can gain insights
into the deviations from the pure Gaussian shape in our models by looking at 
how the code chooses to reduce
the $\chi^2$ of the image. The central shape in the image fits well with component 1 from \S\ref{sec:bb2},
while the extended blobs seen are needed to perfect the $\chi^2$. We believe that the blobs correspond
to extended emissions present in the source, while keeping in mind that the specific shape and positions
of the blobs  mostly reflect the low \textit{uv} coverage of the data set. The contribution of the extended
emission, as measured from the intensities in the image, is about 15\% of the flux of component 1.
\subsection{Channel by channel Gaussian fits}\label{sec:onegauss}
MIDI uses a grism to disperse the two beams into 261 channels before combining them.
  Per channel per time, we have a set of 16 correlated fluxes,
or Fourier amplitudes, one for each baseline.  From these amplitudes a Gaussian
  brightness distribution  corresponding
to infrared emission from this channel's  wavelength  is fitted.  The fitting is done directly in the \textit{uv} plane,
and takes into account only the correlated fluxes. 
 In each wavelength we fitted a single Gaussian, motivated by
the successful  Gaussian fitting of J04 to data  composed of two baselines.
\begin{figure}
\begin{centering}
\includegraphics[width=7cm]{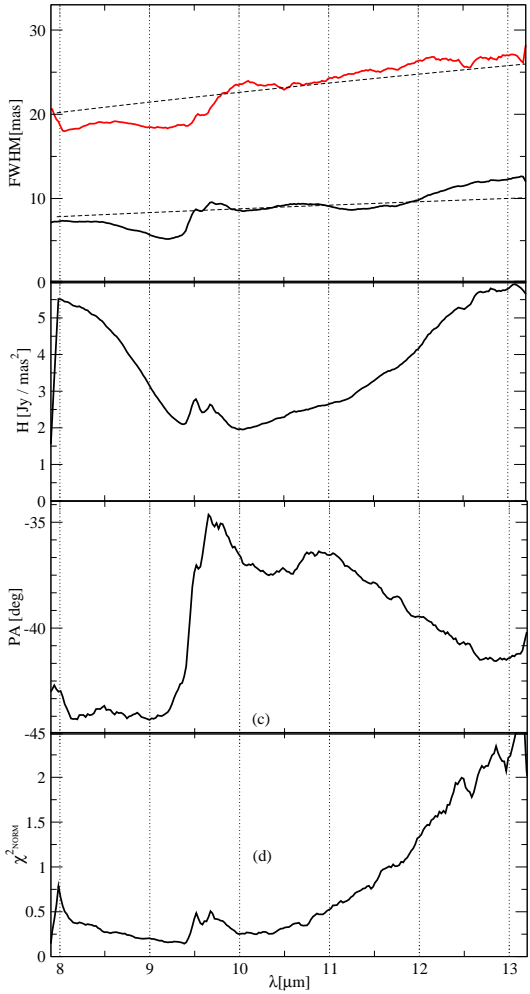}
\caption{Results of one Gaussian fitting. (a) FWHM sizes for the major (red) and minor (black) axes, overlaid
with $r\propto\sqrt{\lambda}$ (dashed);
(b) Gaussian peak; (c) position angle and (d) normalised $\chi^2$, for each wavelength.
 We stress that in this model each  wavelength
was fitted independently, and the  $\chi^2$  in panel (d) is the minimized $\chi^2$ found
for each wavelength.
The `bumps' near 9.7\mum\, are caused by the O$_3$ atmospheric feature.
\label{fig:onegauss}}
\end{centering}
\end{figure}
The fit results are summarized in Fig. \ref{fig:onegauss},  which plots the values of the model's
parameters (i.e.  the major and  minor FWHM sizes, height  and position
angle of the fitted Gaussian) and the resulting  normalised $\chi^2$ for each fit, determined
by dividing the $\chi^2$ by the number of degrees of freedom (16) minus the number of
model parameters (4).
The main results from this approach are:
\begin{enumerate}
\item The FWHM major and minor axis (Fig. \ref{fig:onegauss}a)  lengths increase
 with wavelength. This is expected of centrally
heated dust since longer wavelengths indicate cooler temperatures. and therefore larger radii.  For
optically thin dust in radiation equilibrium where $T\propto r^{-0.5}$, and 
 for blackbody emission where $\lambda$ scales as  $1/T$
(Wien's law),  we get $r\propto\sqrt{\lambda}$ as a crude estimate, which fits the trend seen 
in the fitting of the Gaussian's axis.
The discontinuity
at  $\lambda\simeq9.5$\mum\, is due to the atmospheric 9.7\mum\, Ozone feature and the intrinsic
Silicate absorption.  
\item The Gaussian  flux density (Fig. \ref{fig:onegauss}b) is simply a response to
the varying flux with wavelength.  
\item The Gaussian's position angle  (Fig. \ref{fig:onegauss}c)  is
perhaps the most interesting of results.  First, we fit the same position angle from 8 to 9\mum, followed
by a jump of $\sim$10 degrees, which gradually rotates back but does not return 
to its original angle. The location
of the jump in the position angle is indicative of the wavelength at which the Silicate absorption feature
takes effect. The gradual, linear change in the position can be most easily interpreted
 as asymmetrical absorption,
i.e.  the absorption at 9 micron is more pronounced to the south-west  of the Gaussian  than at 12\mum.
In contrast, a constant PA will indicate a uniform absorption screen is effectively present.  
\item Further, we can deduce
the presence of another dust component from the difference in PA between 8 and 13\mum. The $\chi^2$
(Fig. \ref{fig:onegauss}d)  also supports the presence of an additional component,  
gradually becoming worse from 10\mum\, (with increasing wavelengths).
  Although we cannot determine the geometry of this second component from this method, we can estimate that it
consists of  warm ($\sim$300K) dust, whose blackbody emission begins contributing to the
 flux at $\lambda >$10\mum.
 \end{enumerate}
 
In general, the single Gaussian  per wavelength model fits the data well, and helps to reduce the complex behaviour
of the correlated fluxes  to  a set of four parameters that vary with $\lambda$.  It is, however, a geometrical model
and does not relate the different wavelengths in a physical way. We find it encouraging that the different parameters
are continuous and behave in an `orderly' fashion, although fitted independently.
\subsection{The properties of each component}
We now discuss in  detail the properties of each component in the model,  combining results from
the different modelling methods.
\subsubsection{Component 1}
Component 1 is the dominant source of Mid-infrared emission on MIDI's scales.
 It is  resolved in all of our observations
and its geometrical properties are well constrained to have a
major axis of 20 and a minor axis of 6.4 mas in FWHM with a PA=$-42\deg$.
As Table \ref{tab:fit} shows, and in contrast to the second component,  the temperature and optical depth
are not well constrained. Nevertheless we can establish that  component 1 is composed of hot
 dust (T$\simeq$800K), with a low  scaling factor ($\eta\sim$0.2).
Although we cannot  unambiguously determine the chemical composition of the dust, we observe that it
differs from  the profile of standard interstellar dust. Namely,
the dust should begin absorbing towards 9\mum\, and the optical depth should rise more sharply towards
10\mum,   followed by a shallower  decline up to 14\mum. Figure \ref{fig:onegauss}c
indicates asymmetric absorption toward lower position angles, which may explain the
deviations from a `perfect' fit in both our models for the wavelength range $\lambda>$10\mum.
\textbf{Clumpiness:}
 Evidence for clumpiness comes from the small scaling factor, which is
consistent with  clumpy emission on unresolved scales that effectively reduces the surface brightness
of the source independently of wavelength.
 Attempts to fit the data with models where the scaling factor is
close to unity have not been successful. Extinction in the mid-infrared  towards our line of sight probably
contributes to the low value of $\eta$, but cannot fully account for it  without invoking clumpiness.
More direct evidence for clumpiness comes from the profile of
the correlated flux of \textit{uv} \#4 at 8 \mum, which is   significantly less than our Gaussian model
predicts. This is the observation with the longest baseline and  as such it is  most sensitive
to small scale structure such as clumps. Given a resolution of 5.8 mas at 8\mum\, for that baseline,
we set an upper limit of 0.4pc (5.8 mas) for the size of a clump. Alternatively, one can interpret the missing
flux at that \textit{uv} point as a smooth (rather than clumpy), non-Gaussian flux distribution at sub-parsec scales.
\subsubsection{Component 2}
 Since most of our observations were taken with long baselines, component 2 is mostly over-resolved.
Figure \ref{fig:corrflux2} (filled curves) plots the contribution of this component to the correlated fluxes, illustrating
its presence in only five of the sixteen baselines observed, corresponding to those shorter
than sixty meters in telescope separation.  As a result, its geometrical properties are not well
constrained, while the temperature prediction is confirmed by that found in \S\ref{sec:onegauss}.
 As for the size of the flux distribution, component 2  is an extended Gaussian structure,
with a minor axis of 42 mas and major  axis of 56 mas in FWHM. It's orientation is best fitted
as north-south, with a large uncertainty.
 Component 2  marks the colder, extended part of the torus-like structure, and is
 (together with comp 1) the unresolved source reported by \cite{Poncelet07} and others
(see introduction). Its north-south elongation is common with elongation of the mid-infrared “tongue” 
(Figure \ref{fig:compare}) of NGC\,1068, which  \cite{Bock00} attribute to  
re-emission by dust and UV radiation concentrated in the ionization cone.
\subsubsection{The physical distinction between the components}
In our model the two components are separate and distinct, each with
a fixed temperature. In reality, we expect the dust, on average,  to have a smooth
temperature and density distribution contained in one structure. In this sense,
the two components are an abstraction. However,  our findings here,
and in particular the quality of the fit to the data, indicate that the two-component
 approximation is accurate regarding
the brightness distribution of the dust, and this requires a steeper
temperature gradient than the simple $T\propto\lambda^{-0.5}$
expected for centrally heated, optically thin dust.
\subsection{Summary of modelling  results}
To summarise our main findings, the mid-infrared emission from the core of NGC~1068 within
a beam size of 28 mas ($\sim$2 pc ) is dominated by a warm Gaussian-shaped structure of dust (comp 1)
 with a chemical composition unlike that of dust towards our own galactic centre.
The geometrical properties of this component are independently well constrained  in each of
our methods of investigation. Component 1 is co-linear with, and has similar
size  to the H$_2$O megamaser disk. It is tilted by $\sim45\deg$ to the radio jet, and perhaps
also by a lesser amount ($\sim$15$\deg$) to the ionization cone. The dust temperature is $\sim$800K,
assuming a grey body model. Evidence for clumpiness is mostly indirect, and comes
from the low grey body scaling factor and the deviation from a Gaussian fit for data obtained
 with the longest baseline.
This single Gaussian component cannot account for the total flux as observed with an 8m
telescope, as shown in Figure \ref{fig:totflux}. It also does not fit the data as well for
wavelengths longer that 10\mum\, on our
 shortest baselines, suggesting the need for a second component (comp 2), which is
extended so that it will become over-resolved with baselines longer than 60m.
The filled curves in Figure \ref{fig:corrflux2}, which plots  the contribution of
component 2  to the correlated fluxes, demonstrates how small that contribution is.
As a result, the geometrical properties of this component, including its  Gaussian nature,
are not well constrained.

\subsection{Comparison with previous MIDI studies of NGC~1068}\label{sec:previous}
Two previous studies of the core of NGC~1068 have been made using MIDI.
 Both studies were based on the same set of data: two \textit{uv}-points with baseline
lengths of 42 and 72 meters, obtained during the science demonstration time of the
 instrument. They were taken with the PRISM mode, which offers a low spectral resolution of
$\lambda /\delta\lambda\simeq$30.
The first study, by Jaffe et al. (J04), used a simpler version of the two Gaussian
grey-body components used in this work (\S\ref{sec:bb2}), and their findings
are summarised in the introduction. In general, our
findings here agree with those of J04, deviating
 from their results only in one parameter: the orientation of the extended
component (component 2), which was then reported to be elongated east-west.
This discrepancy is not surprising in light of our finding here that the extended
component is over-resolved in observations with baseline lengths longer than 60m.
This leaves just one observation (in J04) where  component 2  contributes significantly to the
correlated flux, not enough to determine its orientation. The FWHM sizes of the
extended component in Jaffe et al. (30 and 49 mas) are smaller than the sizes reported
here, and so the component was able to contribute to the correlated flux obtained
with the longer baseline (72m), and an estimation of its orientation and axis ratio
could be made.
The second study, by \cite{Poncelet06} (hereafter P06), used a different approach to modelling the
same data set. Instead of Gaussian components, P06 used a combination
of two uniform, circular  disk components  and visibilities rather than correlated
fluxes as the quantity to be fitted. The derived angular sizes and temperatures of the two
disk components are $\sim$35 and 83 mas, and $\sim$361K and 226K, respectively.
Both approaches provided good fits to the data from the two short baselines. This is because
 the differences between a Gaussian model and a uniform disk is small  for short  baselines. 
With the addition of data taken at longer baselines, we can now determine
that the circular disk model of P06 is inconsistent with the MIDI data presented here. 
This is mostly due to the sharp edge of the circular disk model: 
the visibility (and the correlated flux) of a circular disk is proportional to
the absolute value of the J-Bessel function of the first order, which goes down to zero at 
specific baselines depending on the angular size of the disk and the baselines length in 
units of the observed wavelength. For the disk sizes of the P06 model, these zeroes fall between 
baseline lengths of  60m (at 8\mum) to 100m (at 12.8\mum), a range which  is well 
sampled by our observations.  Yet, no trace of the  drop in the correlated flux 
predicted  by the uniform disk model is seen in any of our observations. The zeroes 
in the visibility function appear
for any flux distribution that possesses  a sharp edge. Thus, our data is inconsistent 
with any such flux distribution.  

\begin{figure}
\includegraphics[width=7cm]{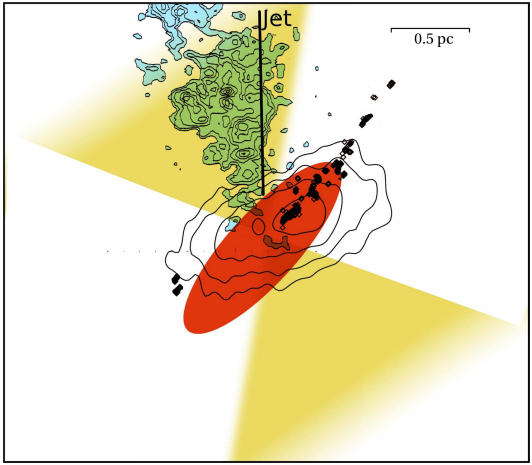}
\caption{Picture summarising the multi-wavelength structures on parsec-scales in the nucleus of NGC\,1068.
 The FWHM of the compact dust (component 1) is sketched in red, centred around the H$_{\,2}$O maser
spots and 5GHz radio emission, both from \citet{Gallimore04}.
Contour levels below the FWHM level have been removed to allow a better comparison between the
radio and the mid-infrared. The ionization cones  inferred from spectroscopy by \citet{Das06}  are shown in yellow,
and the HST [OIII] (\textbf{reduced in scale by a factor $\sim100$}) image contours \citep{Evans91} are shown in blue to indicate the visual orientation of the cone.  \label{fig:sketch}}
\end{figure}
\section{discussion}\label{sec:discussion}
NGC~1068 is considered as the prototypical Seyfert II galaxy. 
It is also the first Seyfert II with broad emission lines seen in polarized, 
scattered light \citep{Antonucci85}.
These observations proved that NGC~1068 harbours an obscured Seyfert I nucleus,
and led to the AGN unification model, at least in its most simple  form,
uniting Seyfert I \& II nuclei.  This is \textbf{the} object for which the 
obscuring  dusty torus was originally invented. The previous sections have
described the {\it appearance} of this structure at mid-infrared wavelengths
and sub-parsec scales, and its geometric relation to structures seen at other wavelengths.

The literature on the spectrum of NGC~1068, 
both observational and theoretical, is very extensive.
Many authors have modelled the production of the overall SED of the galaxy 
within the context of the unified theories.  Without detailed structural observations, 
however, they have found the problem underconstrained.  The early interferometric
work of J04 and W04 led modellers of the emitted spectrum in the direction of collections
of optically thick inhomogeneous (``clumpy") systems.
Following the pioneering work of  \cite{Nenkova02} on clumpy models, 
\cite{Honig06} and \cite{Schartmann08} showed that 
three-dimensional radiative transfer models of clumpy tori were consistent
with the low spatial resolution SED data and the J04 interferometric data of NGC~1068. 
The \cite{Honig06} model of a torus with an inclination of $i=55\deg$ is 
also able to fit the K-band measurements of W04 to a factor of 2 in correlated flux. 
In a later work (\cite{Honig07}) the 
authors have revised the inclination angle of the torus to  $i=70\deg$, 
this time providing excellent fits to both the SED and the previous interferometric 
N-band data, though still falling short of reproducing the K-band interferometric 
data. Most recently, a model with an edge-on torus ($i=90\deg$) was presented 
(\cite{Honig08}). This model was fitted to the SED and one of the interferometric 
data points of J04, reproducing a good fit to the former, and a fit correct to a 
factor of $\sim$2 for the latter. Their most important conclusion is that the 
random cloud arrangement has a significant effect on the SED and images, 
providing remarkably different SEDs for different  cloud arrangements 
and for the same set of large-scale torus parameters.  This accounts for the difficulty in determining the inclination angle. 
 
The observational material presented here is considerably more detailed
than that which was available to the authors cited above, and the results
of their simulations of the production and redistribution of the
radiative energy are generally not presented in a form that can be directly
compared to our measured correlated fluxes as a function of wavelength and
baseline and position angle.  Hence, while these models
seem globally consistent with our present data, we refrain from 
considering whether any of them do or do not {\it fit} it in detail.
Rather we focus on the physical structures we observe and their
relations to the surrounding AGN components.  We also consider
the similarities and differences of the results for NGC~1068 with
those of the nearby Circinus Seyfert II galaxy, \citep{Tristram07}, and
compare the implications of our N-band data with those measured on the
same galaxy by W04 in the K-band. 

\subsection{The inclination and thickness of the dust}\label{sec:inc}
In this section we argue that the nuclear dust structure in NGC~1068 
is circum-nuclear and inclined edge-on, and therefore its 
observed size and the  3:1  ratio of its major/minor axis
is indicative of the  thickness of its inner part rather 
than a projection effect of an inclined disk.
Our main argument here is  that the dust and the maser disk are co-spatial, and since the
 masers are seen edge-on, so is the dust. 
The possibility of detecting the hot dust through the cold dust that surrounds
it has been demonstrated by  several recent three dimensional radiative transfer calculation of tori: \\
\cite{Schartmann05} find that  ``in a homogeneous dust distribution the observed mid-infrared 
emission is dominated by the inner funnel of the torus, even when observing along
the equatorial plane''. Similar results are also found by \citet{Honig06} and 
discussed in more detail in section \ref{sec:previous}. 
We show  that this scenario of an edge-on
geometrically thick torus not only agrees with 3D  radiative 
transfer models, but also is
consistent with other observed structures on milliarcseconds scales:\\
\textbf{(a)} The position angle of component 1 is virtually 
the same as the PA of the western maser spots. 
This can hardly be a coincidence. The theoretical relationship
between water maser excitation and warm molecular dust has been well established
(Neufeld et al. 1994, for example), and the masers are indeed 
expected to be embedded in warm ($>$600K) molecular dust.\\
\textbf{(b)} Apart from the position angle,  the inner radius of the maser disk
is approximately the radius at which dust sublimes \citep{Greenhill98}.
Dust reverberation modelling for the maser disk (Gallimore et al 2001, Fig. 9) 
has established the geometry
of the masers, with those closest to the nucleus outlining a ring 0.6 pc in radius.
The diameter of the ring (1.2 pc) is similar to the FWHM size of the 
major axis of our component 1
(1.35 pc), supporting the conclusion that the masers and the hot dust 
component are co-spatial. \\
\subsection{The orientation with respect to the jet}
Although there seems to be little or no correlation between the relative angle of
 jets and the dust residing in the  galactic disk for  Seyfert galaxies \citep{Kinney00},  
 the naive expectation is that the dust directly orbiting the black hole and the accretion disk is
perpendicular  to the jet.
It is the common belief that the orientation of the jet is closely coupled to the 
spin axis of the black hole, which in turn is affected by the sum 
of all the angular momenta of the material accreted throughout its lifetime.
\cite{Capetti99} estimate the lifetime of the jet as $<$1.5$\times10^5$ years, 
a much shorter time compared to the 
lifetime of the black hole. The current phase of the AGN is not likely to have affected
the spin axis of the black hole, and it is not surprising that the dust and the maser disk
are not aligned with the jet. 
Still, it is interesting to consider the origin of the orientation of the dust 
component and the maser disk, since it seems related neither to the jet nor to 
the galactic plane. We now consider two mechanisms:\\
\textbf{self warping.}  The sub-Keplerian velocities of the masers have been interpreted 
as evidence that the disk is supermassive and will warp due to self-irradiation \citep{Lodato03}. 
This mechanism, however,  only applies to thin structures and therefore cannot
account for the orientation of the dust.  \\
\textbf{misaligned inflow}.  
In this scenario the matter falls misaligned with respect to the 
current orientation of the jet. This can be the result of either matter infalling from
outside the plane of the galaxy, or due to the influence of a torque, perhaps from a 
non-spherical nuclear star cluster. 
 The presence of such a  potential will be most 
likely due to a past minor merger, for which some evidence exists \citep{GarciaLorenzo97}. 
Observations so far have determined that a 
dynamical mass of $6.5\times10^8M_\odot$  is present inside a 50 pc radius
 \citep{Thatte97},  but due to obscuration by dust there is no  information on the
gravitational potential at smaller scales.
This mass is about 12 times the mass of the black hole \citep{Lodato03}, and at a distance a
 few parsecs the black hole's potential still dominates over the potential 
of the star cluster. 
The influence of a star cluster may also explain the non-Keplarian velocity distribution of the masers. 
Finally, the generally linear, stable shape of the jet excludes  non-continuous 
capture of material into the black hole such as individual stars or individual 
dust clouds, which may take an arbitrary route.
To conclude, the identical orientation of both the thin maser disk and the thick 
dusty torus strongly indicate that  it is the  angular momentum of the infalling matter
combined with the gravitational potential at the nucleus that are responsible 
for the current orientation of the maser (i.e. accretion) disk and the inner part of the 
dusty torus that surrounds it. 
When discussing the orientation of the central dust component and the orientations of
 other components related to the nucleus, it is interesting to point out  that \cite{Galliano03}
 have proposed to explain the asymmetrical appearance of the H$_2$ brightness
 distribution (on arcsecond scales) by suggesting that the Compton thick material 
located in the very central parts of the AGN is tilted with respect to the large scale 
molecular gas distribution with an orientation of 30 degrees, which is comparable to our findings.
\subsection{Relation between the torus and the ionization cone}\label{sec:cone}
According to the AGN unification scheme, it is the obscuring torus that is 
responsible for the conical shape of the ionized gas in the narrow line region since it 
blocks the UV radiation from ionizing the material outside the cone. 
The cone's opening angle and orientation are then related to the orientation 
and geometry of the torus, and in particular to the arrangement of the clouds in its 
inner edge.
Now that the inner part of the torus has been resolved, we can compare the two 
quantities and relate them to the predictions of the unification model.
The position angle of the dust is well constrained to be  $-42^\circ$  (see Table 
\ref{tab:fit}). We then expect the ionization cone to be centred perpendicularly
 (i.e. at PA=$48^\circ$).  
The opening angle in this scenario depends mainly on the inner geometry of the torus,
as each cloud in itself is optically thick to UV and X-rays, and trivial geometrical
 considerations limit the cone's angular extent to within
 $\pm90\deg$ of the  torus axis (i.e. -42$<$PA$<$138).
 
 The ionization cone, as seen in HST (Hubble Space Telescope)
  images,  is centred around PA=$10^\circ$, with a small opening angle of 
$\sim45\deg$ (see Evans et al. 1991, for example). As we can see in  Figure
 \ref{fig:sketch}, its axis of symmetry is not aligned with the axis of component 1, as it
lies about 40 degrees further to the north. Yet, due to its small opening angle, the 
cone's angular extent is still within the bounds listed above.  
This discrepancy may be resolved by looking at the kinematics of the outflowing gas.
\citet{Das06} have applied  bi-conical outflow models  to high-resolution
long-slit spectra of the  narrow-line region obtained with the Space Telescope 
 Imaging Spectrograph aboard the HST. These models require the cone to have 
  a wider opening angle, $\sim80\deg$, centred around PA=$30\deg$, rejecting 
  models with PA=$10\deg$ (Das, {\it private comm.}),
and in  agreement with the orientation of our component 1.
Figure \ref{fig:sketch} plots the HST image on top of the kinematic model. 
 The northern edge of the cone
coincides in both the model and the image, and it appears as if only half of the bi-conical outflow
 described by the model is seen in the HST image. The UV radiation which ionizes the gas in the cone
is partially blocked by the obscuring dust closest to the AGN   and is therefore  sensitive to the
configuration of the obscuring clouds. It is not unreasonable to suggest that some of the UV radiation
is blocked by a cloud of dust so that only a part of the gas in the cone is ionize, as seen
in the HST images.
Assuming this, the orientation of the dust component and the ionization cone roughly  match
our expectations and give further support to our interpretation of  component 1 being the
inner hot funnel of the torus-like obscuring object. This implies irregularities in the dust
distribution  and supports the idea of a clumpy dust distribution. 

If this scenario holds then we expect a difference between the appearance of the cone
 in the optical and in the infrared.  The infrared cone (i.e. narrow line region - NLR) should appear 
 more symmetrical with respect to the dust since its appearance does not depend
 on the ionization of its material. The 12.8μm image of \cite{Galliano05} shows 
 extended emission that is quite symmetrical to the north and the south of the 
 central engine, although the northern emission is considerably more pronounced. 
 The emission is centred on PA=$\sim$27 degrees, which is more symmetrical with 
 respect to the dust, with a relatively small opening angle of $\sim$50 degrees. 
 Alternatively, \cite{Poncelet08} study the gas dynamic inside the NLR by looking at 
 several mid-infrared emission lines by extracting spectra along two slit positions. They find 
 the infrared cone to be centred on $PA=10\deg$ with an opening angle of $\sim$80$\deg$.  
 It seems then that in the infrared there is also disagreement between different methods of 
 investigation. A case against this scenario of partial obscuration comes from
 the well defined shape of the cone  over a scale of a few hundred pc, 
 along with the linear shape of the radio jet  both  suggest that these
  irregularities in the dust distribution are stable over $\sim$1500 years. 
 
 To conclude, observations at infrared vs. optical wavelengths, and emission
  line analysis vs. imaging all disagree on the properties of the ionization cone / NLR.
   Further investigation is needed in order to clarify this puzzling aspect.


\subsection{The dust and the radio emission}\label{sec:5Ghz}
Radio observations at  5 and 8.4GHz  of the nucleus of 
NGC\,1068  show a parsec-sized component (hereafter: `radio component'), most likely indicating 
free-free emission from the X-ray irradiated accretion disk \citep{Gallimore04}.
If so, we expect the radio component to have similar geometrical characteristics to 
our component 1. Figure \ref{fig:sketch} shows the 5GHZ contours 
(up to the FWHM level) against the FWHM of component 1. The radio component is 
`thicker' with a major:minor ratio of $\sim2:1$, as opposed to $3:1$ for
component 1 with its  PA of $-57$ degrees.   
The two components are indeed similar in size and orientation.
\subsection{Comparison with Circinus}\label{sec:circinus}
It is interesting to compare our findings with those for the nucleus of the
Circinus galaxy, the only other active galactic nucleus for which the central
mid-infrared source was studied in similar detail. The Circinus galaxy
is similarly regarded as  one of the typical representatives of a Seyfert II galaxy population. It
shows many of the characteristic attributes of Seyfert II sources, such as
narrow and broad emission lines \citep{Oliva94, Oliva98}, an ionization cone
\citep{Veilleux97}, as well as a $0.4\,\textrm{pc}$ edge-on, warped disk of
H$_2$O masers \citep{Greenhill03}. At about $4\,\textrm{Mpc}$,
 the Circinus galaxy is located significantly closer to us
than NGC\,1068. However, due to its lower nuclear luminosity,
$L_{\mathrm{acc}}=10^{10}\,\mathrm{L}_{\odot}$ \citep{Tristram07}, it appears
slightly fainter in the mid-infrared than NGC\,1068.
Using an approach to analyse the MIDI data similar to the one used by us for
NGC\,1068, \citet{Tristram07}  have found the dust in the
nucleus of the Circinus galaxy to be distributed
in two components: (1) a dense, warm
($\sim$330K) component of
$0.4\,\textrm{pc}$ size and (2) a slightly cooler ($\sim$300K),
geometrically thick torus component with a size of $2.0\,\textrm{pc}$. 
The model fits to the data were improved by introducing clumpy-like perturbations in 
the flux distribution, mainly of the larger component, thereby providing arguably the first 
direct evidence of a clumpy structure in an AGN torus.

 As for
NGC\,1068, the compact component coincides with the nuclear water maser disk
in orientation and size and there is evidence for a clumpy or filamentary
distribution of the dust.
Hence there  seems to be some similarity between the dust distributions in these
two Seyfert galaxies.  Especially  striking is the agreement
between the sizes when considering that the torus size is expected to scale
with the square root of the luminosity of the central energy source, and the identical
orientations of the dust and the masers. However, when the properties of the dust 
distributions are compared in more detail, there are several aspects in which 
the dust distribution in NGC\,1068 deviates from that of the Circinus nucleus
significantly.
In the first place, while the temperatures of the extended
components in the two galaxies are comparable, the compact component
in the Circinus galaxy is significantly cooler ($T = 330\,\textrm{K}$) than
the one in NGC\,1068 ($T =800\,\textrm{K}$). The shallow temperature decrease
 of the dust in the Circinus galaxy was attributed to a high degree of clumpiness
 in the torus \citep{Schartmann08}.
In fact, this torus seems to be lacking a significant amount of truly hot
dust. And indeed, we are observing different physical dust structures: in
NGC\,1068 the compact component is interpreted as the dust in the inner
funnel of the torus, which is heated up to near sublimation
temperature.  In the Circinus  galaxy, on the other hand,
the compact component is interpreted as a disk-like dust structure in the centre 
(or `midplane') of the otherwise geometrically thick torus. This structure is  co-spatial 
with the rotating disk of maser emitters. 
As a second difference, in the Circinus nucleus, there is less silicate  absorption in the
 compact component than in the extended component. This was interpreted to be due
  to the silicate  feature in emission, as is expected from optically thin dust in the inner 
  regions of the torus. In NGC\,1068 the inverse is the case, the absorption feature 
  deepens for the compact component.
This, and the unusual spectral characteristics of the dust in NGC\,1068, as opposed
 to the dust in the Circinus galaxy that was found to fit that of standard galactic dust, 
 suggest the reprocessing of dust in NGC\,1068, in analogy to the evolution of dust in protoplanetary disks.
 Several underlying properties can have a major influence on
the observed temperatures and absorption depths: among them are the total luminosity of 
the central energy source, the exact inclination angle of the torus, and the volume filling 
factor of the torus. Proper radiative transfer calculations for the dust distributions in these
 two Seyfert galaxies are needed to bring more light
to this issue.
The relationship between the inner dust component  and the maser disk found in both galaxies is
 the most striking of the features both objects have in common, and it may be that this is indeed
  the generic case in all Seyfert nuclei that also possess maser disks.
To conclude, the existence of a thick configuration of obscuring dust was 
confirmed in both Circinus and NGC\,1068. Both dust configuration are
 found to fit a two component Gaussian model with differences mostly
  in the temperature of the components, and both components
  are similarly related to the water maser disk and the ionization cone.

\subsection{Comparison with other infrared interferometric measurements}\label{sec:Kcompare} 
This galaxy has been observed previously in the infrared with interferometric
resolution in the N-band by us (\cite{Jaffe04a}) and
in the K-band by Wittkowski et al. with the VINCI instrument, (\cite{Wittkowski04} W04), 
and in speckle mode by \cite{Weigelt04}.
These last observations were restricted to effective baselines shorter
than 6 meters by the telescope used, corresponding to $\sim24$ meters at 
our shortest baseline at 8 \mum\, wavelength.  Thus they are primarily 
sensitive to structures larger than any observed here, and we
will not discuss them further.  Here we summarize the 
findings of W04 and discuss
possible nuclear structures that are consistent with 
both their observations and
those reported here. 

W04 found a squared visibility of 0.16$\pm 0.04$ at a projected baseline 
length of 45.8 meters and position angle of 44.5 degrees, corresponding
to a resolution of $\sim$5 mas.  The authors favour a multi-component 
model where 52 mJy originates in an unresolved source of 
size $\lesssim$5 mas and 75 mJy arises on scales of the order 
of 40 mas or larger. Again this larger component
will not be discussed here.  The smaller emission was attributed to
thermal emission from hot (1000-1500 K)  dust at the 
inner cavity of the torus, with a
possible contribution of direct  plus scattered light from the central engine.  
The resolution achieved by VINCI at 2.2 \mum\, is only $\sim 30$\% smaller than 
that we reach on our longest baseline, 129 meters at our shortest 
wavelength, 8 \mum,
thus in spatial terms the measurements can be compared directly.

The interpretation of these measurements is clouded by several
uncertainties.  First, the interferometric observations provide no
direct positional information.  Thus we do not know the position 
of the small K-band component relative to the galactic nucleus
or relative to the extended features seen with MIDI, except that
it must lie within the 56 mas field of view of the VINCI fibers,
presumably centered on the nucleus.  Second, the extinction
in front of the K-band component is uncertain.
If this component is seen through the same extinction as our Component 1
(Table 2), then the K-band absorption may be estimated by scaling
the depth of the silicate absorption by a standard extinction curve 
(c.f. \cite{Schartmann05}, Fig. 3), 
yielding an optical depth of $\tau_K \sim 3$ but with large
uncertainties, particularly if the dust is of non-standard composition.
If the K-band component is not at the nucleus, e.g. represents a
clump of warm dust clouds as suggested by W04 and \cite{Honig08}, then
perhaps our line of sight to it does not pass through the main body
of dust absorption, and the extinction could be significantly smaller.
Third, as we have seen in Section 4, the short wavelength, long baseline MIDI
results can be well modelled by a smooth gaussian of size 20$\times$ 6
mas, with no evidence of unresolved emission corresponding to that
found by VINCI.  However, the low scaling factor of Component 1, 
$\alpha = 0.25$ (Table 2), points to the possible existence of substructure 
on size scales below the resolution of MIDI.  
For the 8 \mum\, flux of any of the
sub-components we can then only set an upper limit of $\sim 750$ mJy 
corresponding to the correlated flux seen on a 129 meter baseline (Figure 3,
panel 1). Assuming again a standard extinction law, its intrinsic 8 \mum\, flux
would be $\lesssim 1.3$ Jy. 
 
Under these circumstances, we consider only two
possibilities for the K-band source where reasonable additional
assumptions can be made concerning the above uncertainties:
either emission from the accretion disk around the central black hole, or 
from a small body of hot dust, as discussed by W04 and also proposed by \cite{Honig08}.

\subsubsection{Central Accretion Disk}
In this case we assume that the source lies behind the absorption seen
in N-band and assume that for the intrinsic flux from the accretion
disk S$_\nu(2.2\mu) = 0.05 \times e^{3} =  1 $ Jy and $S_\nu(8\mu)\leq 1.3$ Jy. 
This would correspond to a $\nu^{-0.2}$ spectrum, not unreasonable for an
accretion disk.  
In addition, this accretion disk flux is consistent with the total accretion disk luminosity:
From the total infrared emission $\int \nu S_\nu d\ln\nu\simeq 2.5\ 10^{15}$ Jy Hz 
(estimated from the data assembled in the {\it NASA/IPAC ExtraGalactic Database}) one infers 
the bolometric emission of the disk to be $\nu S_\nu(peak)\sim 7 \times 10^{15}$ Jy Hz (assuming that the UV luminosity exceeds the IR luminosity
by a factor $\sim 3$, because light
emitted out of the torus plane is not reprocessed into the infrared).
At a distance of 14 Mpc this corresponds to $\sim 1.5\ 10^{45}$ erg s$^{-1}$, similar to
other estimates of the bolometric luminosity of NGC~1068. 
\cite{Schartmann05} discuss  likely SEDs for central accretion disks, 
based on observations of Type I AGNs (e.g. Walter et al., 1993,
Zheng et al., 1997) and more modern theoretical models where observations are lacking.
In most cases the spectrum is relatively flat in the optical-IR region,
$-0.5<\alpha_\nu<+0.3$ (where $S_\nu\propto \nu^{\alpha_\nu}$) and breaks sharply
down shortwards of the Lyman edge ($\lambda_{peak}\sim 0.1\mu$m i.e. $\nu_{peak}\sim 3\ 10^{15}$ Hz) 
With this value of $\nu_{peak}$ and the bolometric luminosity 
we find $S_\nu(peak)\sim 2$ Jy. The relatively
flat spectrum thus inferred between $\nu_{peak}$ and 2.2 \mum\, is consistent
with the acceptable range of $\alpha_\nu$.

We conclude, therefore, that the fluxes and limits measured with 
the VLTI at 2 and 8$\,\mu$m are consistent with an accretion disk spectrum.
In fact, if the emission spectrum of the accretion disk and the (relatively low) foreground 
extinction value are correct we {\it expect} to see the accretion disk in the K-band measurements.
\footnote{This is at variance with \cite{Honig08}, who on the basis of a simple theoretical 
$\nu^{1/3}$ accretion disk spectrum peaking at $\nu \sim 10^{17}$ Hz, 
for which little observational evidence exists, argued for an undetectably
low accretion disk flux at $\lambda > 1\mu$m.}

\subsubsection{Dust Cloud Emission}
Several authors (Nenkova et al., 2002, H\"onig 2006, Dullemond et al 2005) have proposed that
AGN dust tori are clumped and H\"onig (2008) suggest that the small 
component of the VINCI observations is a warm dust cloud or 
compact cluster of such clouds.  To evaluate this suggestion in
the light of our observations requires an additional assumption
about the K-band extinction.  If, as above, we assume the clouds
to be near the nucleus this would to be
of order $\gs 3$ mag, the intrinsic K-band flux is $\gs 1$ Jy
and the $8\mu{\rm m}\rightarrow 2.2\mu$m spectral index is $\alpha_\nu \ls -0.2$.
For black-body emission this implies a color temperature $T_{color}> 1300$ K,
which is rather warm for dust, but not excluded, particularly if 
the dust is more refractory than standard silicates.  If, in fact,
the dust clouds are in the immediate vicinity of the nucleus, i.e.
$r<2.5$ mas, their equilibrium temperature derived from the above luminosity would be 
much higher than this, making dust emission very unlikely. However, as discussed at the 
beginning of this section, there is no direct evidence that the emission feature is 
physically so close to the nucleus.  
If the unresolved K-band source is {\it not} centered on the nucleus 
but above the main body of dust clouds, or seen through a hole in
these clouds, then its intrinsic K-band flux could be as low as the
observed 50 mJy, in which case the N-band upper limit provides no useful information
on its properties.

Thus we conclude that the K- and N- band interferometric data can be
straightforwardly be explained as emission from a hot central accretion disk
if reasonable assumptions are made on the foreground emission and the accretion disk spectrum.  The
data can also be explained as emission from a group of hot dust clouds,
but then it is unlikely that these clouds are centered around the central source. Additionally in this case, the extinction towards the core must be
sufficient (i.e. $\tau_K >> 3$) to suppress the expected K-band emission from 
the nuclear accretion disk below the observed 50 mJy level.
\section{Conclusions}\label{sec:conclusions}
In this paper we present interferometric observations of the nucleus
of NGC~1068, using MIDI  at the VLT. Extensive \textit{uv} coverage of
sixteen baselines with a maximal resolution of 7 mas has allowed us to analyse
the mid-infrared (8-13\mum) emission from the obscuring torus in great detail.
We find the measurements consistent with emission from  a compact ($0.45 \times 1.35$ pc)
component with a Gaussian flux distribution that is composed of hot
($\sim$800K) silicate dust with an unusual absorption profile, which we
identify as a funnel of hot dust associated with the obscuring torus.
 The emission is co-linear, and likely co-spatial, with the
well studied H$_2$O megamaser disk, and therefore tilted by $45 \deg$ with respect
to the radio jet. A second, more extended ($ 3 \times 4$ pc) component
of warm ($\sim$300K) silicate dust  is weakly detected, which we identify as `body' of the torus. 
This second component
is mostly over-resolved and its properties are not well constrained.
We discuss the physical origin of the
emission with respect to the torus, the masers, the ionization cone, and the radio jet.
A direct image of the source at 8 \mum,
obtained with maximum entropy image reconstruction, is presented as well.
Since the obscuring torus is a crucial component in the accepted picture of an AGN,
resolving the structure of the torus illuminates many aspects of the AGN picture.
We show that in many aspects the nucleus of NGC\,1068 is irregular: the orientation of the
dust is tilted with respect to the jet; the PA of the visible ionization cone does not match the 
PA of the inner dust funnel; and  the chemical composition of the
dust in the torus is different than dust observed in the Milky Way.
\section*{Acknowledgements}
This research was supported  by the Netherlands
Organisation of Scientific Research (NWO) through
grant 614.000.414.
Based on observations collected at the European Southern
Observatory, Chile, program numbers 076.B-0743, 277.B-5014(B). \\
The authors would like to thank the anonymous referee,  Julian Krolik, Leonard Burtscher
and Demerese Salter for their helpful comments. 

\label{lastpage}
\end{document}